\documentclass[twocolumn, superscriptaddress]{revtex4}

\usepackage{graphicx}
\usepackage{bm}
\usepackage{amsmath,amssymb,amsfonts}
\usepackage{float}
\usepackage{hyperref}

\usepackage{xcolor}
\usepackage[normalem]{ulem}

\begin{document}

\title{Volume-Scaled Common Nearest Neighbor Clustering Algorithm with Free-Energy Hierarchy}

\author{R. Gregor Wei{\ss}}
\author{Benjamin Ries}
\author{Shuzhe Wang}
\author{Sereina Riniker}
\email{sereina.riniker@phys.chem.ethz.ch}
\affiliation{Laboratory of Physical Chemistry, ETH Z{\"u}rich, Vladimir-Prelog-Weg 2, 8093 Z{\"u}rich, Switzerland}

\begin{abstract}
The combination of Markov state modeling (MSM) and molecular dynamics (MD) simulations has been shown in recent years to be a valuable approach to unravel the slow processes of molecular systems with increasing complexity. While the algorithms for intermediate steps in the MSM workflow like featurization and dimensionality reduction have been specifically adapted for MD data sets, conventional clustering methods are generally applied for the discretization step. This work adds to recent efforts to develop specialized density-based clustering algorithms for the Boltzmann-weighted data from MD simulations. We introduce the volume-scaled common nearest neighbor (vs-CNN) clustering that is an adapted version of the common nearest neighbor (CNN) algorithm. A major advantage of the proposed algorithm is that the introduced density-based criterion directly links to a free-energy notion via Boltzmann inversion. Such a free-energy perspective allows for a straightforward hierarchical scheme to identify conformational clusters at different levels of a generally rugged free-energy landscape of complex molecular systems.
\end{abstract}

\maketitle

\section{Introduction}

The interest in Markov state models (MSMs) to gain insights into metastable processes from molecular dynamics (MD) simulations has increased over the last decade~\cite{Chodera2007, Pande2010, Prinz2011, Bowman2013, Chodera2014, Husic2018}. Especially, the determiniation of MSMs for very high-dimensional systems has evolved and proven useful for the understanding of slow dynamics in complex molecular systems. Examples include application to liquids~\cite{Netz2018}, peptide dynamics~\cite{Witek2016}, protein folding~\cite{Lane2011} and design~\cite{Caflisch2007}, ligand binding~\cite{Plattner2015}, RNA fraying~\cite{Pinamonti&Bussi}, and polymer dynamics~\cite{Sidney2005_I, Sidney2005_II}.

The MSM methodology includes advanced algorithmic procedures for featurization~\cite{Scherer2019, Wu2020}, dimensionality reduction~\cite{Schwantes2013, Perez2013}, and discretization~\cite{Macqueen1967, Steinley2007, Kaufman2009, Keller2010, Sittel2016, Rodriguez2014, Huang2017} to assign states that label different attractors of the underlying dynamical system~\cite{Prinz2011, PyEMMA2015}. In particular, the discretization step utilizes clustering algorithms~\cite{Macqueen1967, Steinley2007, Kaufman2009, Keller2010, Sittel2016, Rodriguez2014, Huang2017} to group similar conformations in the Boltzmann-weighted data sets of the complex molecular simulations. Currently, full-partitioning algorithms~\cite{Macqueen1967, Steinley2007, Kaufman2009, Prinz2011, Wehmeyer2018} are still the most commonly used approaches, where all data points are assigned to a cluster. However, the drawback of full-partitioning algorithms is that the resulting discretization may fail to fulfill the assumption of Markovianity because of fast ballistic transitions across cluster boundaries~\cite{Buchete2008, Keller2010, Schuette2011}. 

More recently, core-set MSMs~\cite{Buchete2008, Keller2010, Schuette2011, Lemke2016, Lemke2018, Pinamonti&Bussi} have been introduced to avoid the contributions from ballistic transitions by introducing a transition region (i.e. parts of the trajectories remain unassigned as so-called `noise' points) such that the cluster boundaries are spaced. This means that the core-set clusters identify only the free-energy minima, while the rest of the conformations become `noise' points. Subsequently, a milestoning procedure counts a transition if a trajectory exits a cluster, passes through the intermediate noise region, and enters another cluster~\cite{Buchete2008, Schuette2011}. Thus, time correlations (i.e. memory effects) are diminished in the transition counts such that the Markovianity is retained. To determine the core sets, density-based clustering algorithms~\cite{Lemke2016, Lemke2018, Witek2016, Nagel2019, Huang2017, Pinamonti&Bussi} have proven useful. These algorithms cluster the data points within regions of maximum density and declare those in sparsely sampled regions as noise points. 

A density-based clustering that has recently been used for MSM construction of MD simulations is the common nearest neighbor (CNN) algorithm~\cite{Keller2010, Lemke2016, Lemke2018}. A second method introduced by Sittel and Stock~\cite{Sittel2016} translated the density notion into a free energy-based clustering specifically for MD data, which we will abbreviate as the Sittel-Stock algorithm. In general, both algorithms cluster $m$ data points of a set $X=\{\bm{x}_1, \bm{x}_2, \dots, \bm{x}_m\}$ with $X\subset\mathbb{C}^n$ where the $n$-dimensional data space $\mathbb{C}^n$ can be a reduced conformational space or the full phase space of the dynamic system. The algorithms determine the neighbors of a given data point within a hyperspherical neighborhood of radius $R$. Likewise, $R$ is also denoted as the \textit{cutoff} distance for data points outside the hypersphere. Thus, the neighborhood of point $\bm{x}_i\in X$ can be formalized as 
\begin{equation}
M_i = \left\{ \bm{x}_j \, \bigl| \, \vert\bm{x}_i - \bm{x}_j\vert < R; \, \bm{x}_j \in X \setminus \bm{x}_i \right\} ,
\end{equation}
which defines the number of neighbors of $\bm{x}_i$ by the cardinality $\mathrm{card}(M_i)$. The CNN and the Sittel-Stock clustering determine the similarity of two points using a density-based \textit{similarity} measure, i.e. a point $x_i$ is assigned to a given cluster if (i) there is a point $x_j$ in this cluster within a distance threshold (which can be chosen different from $R$~\cite{Keller2010, Sittel2016}), and (ii) if $M_i$ and $M_j$ fulfill a density-based similarity criterion. The similarity measure of CNN requires the two points $x_i$ and $x_j$ to share $N$ neighbors within the intersection of their hyperspherical neighborhoods, i.e. $\mathrm{card}(M_i \cap M_j) \geq N$. To define the similarity criterion for the Sittel-Stock algorithm, the number of neighbors $\mathrm{card}(M_i)$ of a point $x_i$ is used to determine a free-energy estimate by Boltzmann inversion
\begin{equation}
F_i = -k_\mathrm{B}T \mathrm{ln}\left(\frac{\mathrm{card}(M_i)+1}{N_\mathrm{max}+1}\right) ,
\label{eq:rdb}
\end{equation} 
where $N_\mathrm{max} = \mathrm{max}\{ \mathrm{card}(M_1), \mathrm{card}(M_2), \dots, \mathrm{card}(M_m)\}$ is the maximal cardinality of all neighborhoods. The increment of one in the numerator and denominator inside the logarithm accounts for the fact that a point $\bm{x}_i$ is not contained in its own neighborhood $M_i$. The Sittel-Stock algorithm defines two points $x_i$ and $x_j$ as similar if their free-energy estimates $F_i$ and $F_j$ are smaller than a given threshold $F_\mathrm{thresh}$.

An important difference of the similarity criteria of CNN and Sittel-Stock clustering lead to a crucial memory performance advantage of the latter. The similarity criterion for CNN requires storing all neighborhoods to determine $\mathrm{card}{(M_i \cap M_j)}$. Sittel-Stock clustering only stores the free energy estimate $F_i$. Hence, the memory requirement of CNN will typically exceed that of the Sittel-Stock algorithm.

The free-energy perspective from the Sittel-Stock algorithm  highlights the potential issues when clustering MD data from an intrinsically rugged free-energy landscape with a single cutoff and similarity threshold. As the clustering criteria consider only data points below a certain free-energy threshold, higher free-energy basins may be omitted. However, such higher lying minima could be crucial to detect transition paths in complex molecular systems. The opposite issue arises when choosing a too high threshold, which causes several free-energy basins to be merged into a single cluster. The solution to these issues presented by Sittel and Stock~\cite{Sittel2016} is a hierarchical approach, which generates a tree of clusters at different free-energy based thresholds. Their implementation explores this hierarchical tree in a bottom-up approach, starting from the lowest free-energy threshold (i.e. highest density) that still clusters a minimal portion of the data, and gradually increasing the threshold until all data is clustered. A hierarchical approach was also mentioned in the context of the CNN algorithm,\cite{Lemke2016, Lemke2018} but a robust hierarchical scheme has not been suggested so far. 

CNN has natively been employed for the construction of core-set MSMs~\cite{Witek2016, Lemke2016, Lemke2018}. In contrast, applications of the Sittel-Stock clustering required an additional algorithmic step to include core sets~\cite{Nagel2019}. In particular, after the discretization by the Sittel-Stock algorithm the discrete trajectories and MSM statistics still seemed to include ballistic transition counts across neighboring cluster boundaries on example data of a villin headpiece folding simulation~\cite{Sittel2016, Nagel2019}. Hence, to restore Markovianity in discretized trajectories Nagel et al.~\cite{Nagel2019} employed 'dynamical coring'~\cite{Jain2014} that effectively counts transitions for which the trajectory resides in the new state for a user-defined minimum time, e.g. 3~ns. The Sittel-Stock clustering creates a free-energy hierarchy of clusters, implying a hierarchy in metastability and respective residence time statistics. Hence, dynamical coring with a fixed waiting time distorts the first-passage time statistics of faster living states. Additionally, a simplification of core-set MSM construction by avoiding such additional algorithmic steps is preferable.

In this work, we address the issues mentioned above in the context of the CNN algorithm. We start by translating the CNN algorithm into a free-energy interpretation such that the density-based parameter choice can be interpreted as a free-energy difference. For this, we implement the calculation of the hyperspherical intersection volumes to determine the actual data-point densities for a new \textit{similarity} criterion. The implementation of such a volume-scaled common nearest neighbor (vs-CNN) algorithm rescales the intersection volumes by the full hyperspherical volume such that numeric overflow for large dimensionality can be avoided. The new density-based notion can be used in a free-energy interpretation by Boltzmann inversion such that a hierarchical clustering approach can be conducted. In contrast to the Sittel-Stock algorithm, our hierarchical approach starts in a top-down manner from a high free-energy based threshold, for which all or most points are clustered, and gradually decreases the threshold until no further sub-clusters are identified. The proposed hierarchical scheme is based on threshold parameters, which translate to physically relevant measures such as free-energy differences between hierarchical levels and cluster sizes corresponding to a cumulative time threshold per cluster. First, the algorithm is tested on Brownian dynamics (BD) of a particle in a circular five-well potential, where the wells are at different (free) energy levels. Secondly, we employ the hierarchical vs-CNN approach on a MD trajectory that samples multiple folding and unfolding events of the villin headpiece taken from Ref.~\cite{Piana2012}. The same trajectory has been used to generate MSMs based on the Sittel-Stock clustering. Thus, our results presented here can be compared to Refs~\cite{Sittel2016, Nagel2019}. The vs-CNN algorithm and the associated hierarchical scheme are described in Section~\ref{sec:theory}. The details of the BD simulations, the dimensionality reduction of the MD trajectory, and the MSM construction are provided in Section~\ref{sec:methods}. In Section~\ref{sec:results:bd}, the MSM results of the BD system are discussed, and we demonstrate how MSMs with a discretization based on a single cutoff fail to describe the systems' dynamics correctly. This in turn can be resolved by our hierarchical vs-CNN algorithm. The clustering and MSM results of the villin headpiece data are presented in Section~\ref{sec:results:hp35}. In particular, the sensitivity of the hierarchical vs-CNN approach on parameter settings is depicted and the MSM results are compared for several cases of reduced dimensionality. In Section~\ref{sec:conclusion}, we conclude how the hierarchical vs-CNN algorithm achieves improved core-set identification for proper MSM timescale separation even in large data dimensionality and with a reduction of algorithmic steps. 

\section{Theory}
\label{sec:theory}

\subsection{Volume-Scaled CNN Algorithm}

To translate the CNN algorithm into an explicit density notion, the local data-point density is calculated for the intersection of $M_i \cap M_j$. This density is determined by the cardinality $\mathrm{card}(M_i \cap M_j)$ and the volume of intersection of two hyperspheres with radius $R$ centered at points $x_i$ and $x_j$ in $\mathbb{C}^n$. This intersection volume $I_{n,R}(x_i, x_j)$ is expressed using the regularized incomplete beta function $\mathcal{I}(x,a,b)$ such that~\cite{Li2011}
\begin{equation}
I_{n,R}(x_i, x_j) = V_n(R) \cdot \mathcal{I}\left(1-\frac{|x_i - x_j|^2}{4R^2}, \frac{n-1}{2}, \frac{1}{2}\right) \, ,
\end{equation}
where the full volume of a hypersphere grows with respect to the dimensionality $n$,
\begin{equation}
V_n(R) = \frac{\pi^{n/2}}{\Gamma\left(\frac{n}{2} + 1 \right)} R^n\, ,
\end{equation}
which in turn is expressed by the $\Gamma$-function. The volume of a sphere and thus the intersection of two spheres increases with $n$, which limits the computation to an upper bound in $n$ due to numeric overflow. However, the volume computation can be rescaled to $V_n(R) = 1$ such that the rescaled intersection volume $I_{n,R}(x_i, x_j) / V_n(R)$ is smaller or equal to one.

Thus, the CNN algorithm can be directly translated into the vs-CNN algorithm, where two points $x_i$ and $x_j$ are \textit{similar} when the point density within the intersection of their neighborhoods is above the rescaled density threshold $N/V_n(R) = N$ such that
\begin{equation}
\frac{\mathrm{card}(M_i \cap M_j)+2}{\mathcal{I}\left(1-\frac{|x_i - x_j|^2}{4R^2}, \frac{n-1}{2}, \frac{1}{2}\right)} \geq N \,.
\label{eq:cd}
\end{equation}
Again, the increment of two in the numerator accounts for the fact that the points $\bm{x}_i$ and $\bm{x}_j$ are not contained in their respective neighborhoods $M_i$ and $M_j$. Therefore, to keep the complete algorithm simple, two points $\bm{x}_i$ and $\bm{x}_j$ are clustered if (i) they are within the cutoff distance $\vert \bm{x}_i - \bm{x}_j \vert < R$ and if (ii) their neighborhoods $M_i$ and $M_j$ fullfill the similarity criterion in Eq.~\ref{eq:cd}.

$R$ and $N$ remain as free parameters in the vs-CNN algorithm. In particular, by increasing $R$ and/or decreasing $N$ the density criterion is lowered and \textit{vice versa}. Moreover, the choice of $R$ and $N$ strongly depends on the overall data-point density, which in turn is influenced by an interplay between sampling and dimensionality. For instance, for systems with a large number of dimensions the parameter choice has to consider rather sparse sampling. Additionally, for computational and memory efficiency a large data set may be sliced, i.e. every $m$th frame is taken. The choice of the slice frequency $m$ again affects the choice of $R$ and $N$. In case of the CNN algorithm, no rigorous evaluation of the choice of the free parameters has been reported so far in the literature. Therefore, applications of CNN clustering include a grid search on $R$ and $N$~\cite{Lemke2016, Lemke2018}.

\begin{figure}[t]
\centering
\includegraphics{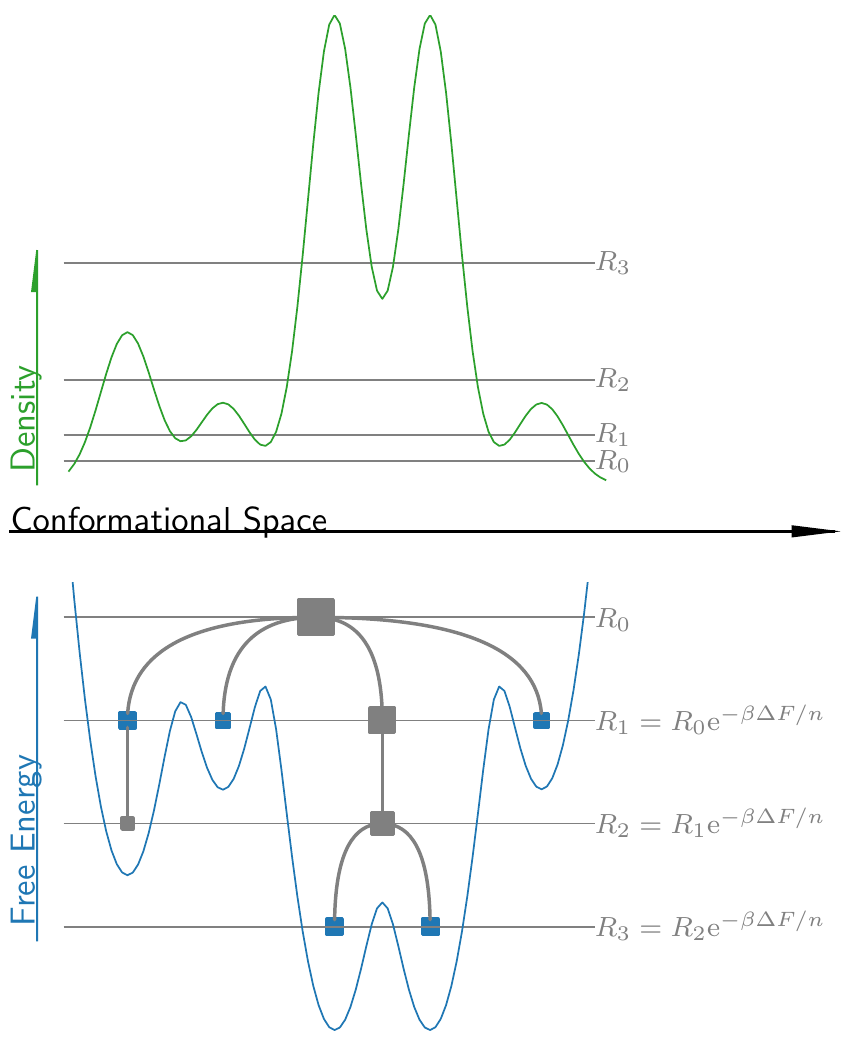}
\caption{Simple example of a density profile in green (top) and its Boltzmann inversion (i.e. free-energy profile) in blue (bottom) to showcase that the density peaks (i.e. free-energy basins) cannot be identified with a single cutoff represented by gray lines at cutoffs $R_3 < R_2 < R_1 < R_0$. In the free-energy profile, the hierarchical tree is drawn with squares marking the cluster centers and relative size. The recursive relations are explicitly denoted for the respective hierarchical level. The blue colored clusters would be the final output.}
\label{fig1}
\end{figure}

\subsection{Free-Energy Hierarchy}

Using a grid search to determine the single values of the free parameters of CNN clustering may not be able to resolve a free-energy landscape correctly. The example density profile in the top panel of Figure~\ref{fig1} illustrates how the five density peaks cannot be distinguished by with a single density threshold, e.g. cutoffs $R_3<R_2<R_1<R_0$. The smallest cutoff $R_3$ identifies only the two largest peaks of the profile and three out of five metastable states are assigned to noise. The cutoff $R_2$ identifies one of the missing metastable states but the two largest density peaks cannot be distinguished anymore. Merging of metastable states occurs also with an even lower cutoff $R_1$, while the smallest density peaks are detected. 

The corresponding free-energy profile (i.e. the Boltzmann inverse) in the bottom panel of Figure~\ref{fig1} shows the combination of the different cutoffs as a hierarchical tree. It starts by finding a sparse density criterion (i.e. cutoff $R_0$) at fixed $N$, for which most data points are comprised in a single cluster. Next, the hierarchical tree is iteratively explored for decreasing cutoffs,
\begin{equation}
R_{i+1} = R_{i}\cdot \mathrm{e}^{-\beta\Delta F/n}\, ,
\label{eq:hierarchy}
\end{equation}
where $\Delta F$ is a free-energy difference in units of thermal energies, i.e. $\beta^{-1} = k_\mathrm{B}T$ with $k_\mathrm{B}$ being the Boltzmann constant, and $T$ the absolute temperature. Thus, at the hierarchical level $i+1$ the clusters from level $i$ are re-evaluated for the decreased cutoff $R_{i+1}$. Moreover, only if a cluster from step $i$ yields two or more new clusters at step $i+1$, the new clusters replace the cluster from step $i$. This requirement prevents simple shrinking of the clusters. Note that the new free-energy based parameter $\Delta F$ replaces $R$ and $N$ as free parameters and provides an interpretability of the clustering.

To avoid overdiscretization, two additional time-based parameters are introduced. The first parameter is the minimal size of a cluster to keep $N_\mathrm{keep}$, which was already introduced in the original CNN algorithm~\cite{Keller2010, Lemke2016, Lemke2018}. The second parameter is the minimal size of a cluster to split $N_\mathrm{split}$ in the hierarchical approach going from step $i$ to step $i+1$, given that $R_{i+1}$ introduces at least two clusters of size $N_\mathrm{keep}$. Both parameters can be chosen by interpreting the sum of frames inside a cluster as an accumulated time content, i.e. the product of the number of frames and time step size $\Delta t$. For example, a cluster containing a frame count accumulating to a total time larger than $N_\mathrm{split} \cdot \Delta t=$10~ns is only re-evaluated for sub-clusters that have at least $N_\mathrm{keep} \cdot \Delta t=$1~ns of aggregate time.

In principle, any recursive relation between $R_{i+1}$ and $R_i$ can be applied. For instance, Lemke and Keller~\cite{Keller2010, Lemke2016, Lemke2018} used a uniform cutoff sequence such that $R_{i+1} = R_i - \Delta R$, with a chosen step size $\Delta R$. However, the choice of $\Delta R$ is as arbitrary as that of $R$ and $N$. This prevents an direct interpretability, and thus introduces one more free parameter. More importantly, a free-energy based hierarchy is ideally suited for molecular systems because the underlying data is Boltzmann distributed. The illustration in Figure~\ref{fig1} demonstrates that the thresholds $R_0$ -- $R_3$ are equidistant in free energy, but exponentially distributed in terms of density. This means that the step size in the density cutoff is large at the top of the hierarchy, and becomes automatically smaller towards the bottom. Such a cutoff scaling reduces the risks of either a too fine-grained hierarchical tree at the top or overlooking the (possibly most important) metastable states at the bottom.

\section{Methods}
\label{sec:methods}

\subsection{Brownian Particle in a Circular Five-Well Potential}

\begin{figure}[t]
\centering\includegraphics{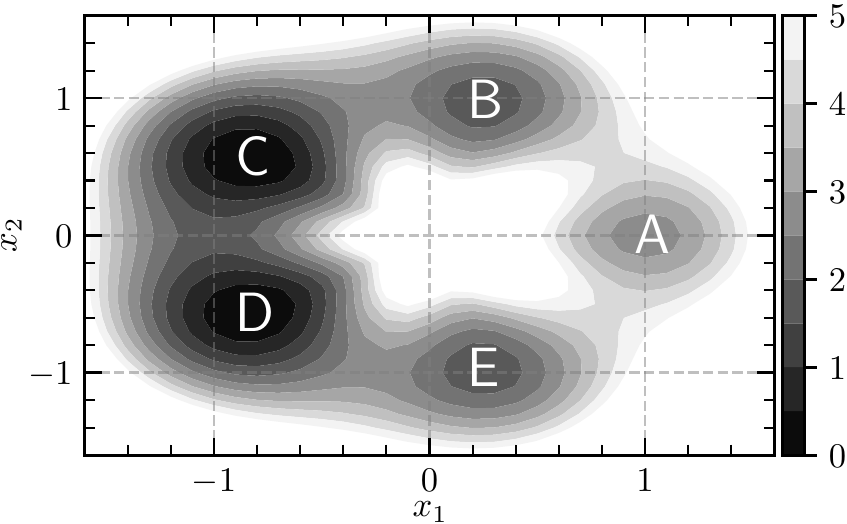}
\caption{Two-dimensional potential (Eq.~\eqref{eq:potential}) with five minima and a bias towards decreasing values of $x_1$. The minima are labelled with A -- E for later reference.}
\label{fig2}
\end{figure}

As an illustrative dynamic test system, we considered a Brownian particle in a two-dimensional potential $V(\vec{x})$ using the overdamped Langevin equation,
\begin{equation}
\dot{\vec{x}} = -\frac{D}{k_\mathrm{B} T} \vec{\nabla} V(\vec{x}) + \sqrt{2D}\vec{R}(t)\, ,
\label{eqBD}
\end{equation}
where $\vec{R}(t)$ is Gaussian random noise with zero mean and variance $\langle \vec{R}(t) \vec{R}(t')\rangle = \delta(t-t')$. The system is presented in units of the Brownian timescale $\tau_B = \lambda_B^2/D$, where $\lambda_B=D=1$ are the Brownian length scale and the diffusion constant. To showcase the hierarchical approach for the vs-CNN clustering, a circular potential was applied,
\begin{align}
V(\vec{x}) = & \mathrm{cos}\left( k\cdot\mathrm{arctan2}(x_2, x_1) \right) \nonumber \\
 & + 10\left( \sqrt{x_1^2 + x_2^2} - 1 \right)^2 \nonumber \\
 & + \frac{c}{\sqrt{1 + x_1^2/x_2^2}} .
\label{eq:potential}
\end{align}
Five minima were obtained by the multiplicity $k$~=~5 and a bias towards decreasing values of $x_1$ is achieved by $c$~=~1. Figure~\ref{fig2} labels the five energy basins A, B, C, D, and E, which are spread over several thermal energies. The Langevin equation (Eq.~\ref{eqBD}) was numerically integrated using the Ermak-McCammon algorithm~\cite{Ermak&McCammon} in a single trajectory of length $10^4~\tau_B$ using a time step of $\delta t = 10^{-3}~\tau_B$ and a writing frequency of 20 steps.

\subsection{MD Simulation}

A 305~$\mu$s long all-atom MD trajectory of the Nle/Nle-mutant of the villin headpiece was used as a  realistic test system, taken from the repository of D. E. Shaw Research (DESRES)~\cite{Piana2012}. The simulation contained a protein consisting of 35 amino acids in explicit water and was performed on the Anton supercomputer using the Amber ff99SB*-ILDN force field~\cite{ILDN}. The temperature was kept at 360~K and the amino acid coordinates were stored every 200~ps.

\subsection{Dimensionality Reduction}

The villin headpiece coordinate space was reduced to the backbone dihedral angles of the protein. We used the same dimensionality reduction as in Ref.~\cite{Nagel2019}, which presented Markov state models of the same trajectory using the Sittel-Stock clustering with dynamical coring. The dihedral angle principal component analysis with maximal gap shifting (dPCA+)~\cite{Sittel2017} was used to reduce the dimensionality of the backbone dihedral angles. Details and illustrations of the villin headpiece principal component (PC) space can be found in Refs~\cite{Sittel2016, Sittel2017, Nagel2019}. As in Ref.~\cite{Nagel2019}, we chose the first five and the seventh PC as reduced conformational space. This choice by Nagel et al. was based on the visual detection of basins in the free-energy surface. In addition to this six-dimensional PC space, we employed the hierarchical vs-CNN approach on the more agnostic approach of taking the first 15, 20, or 30 dimensions.

\subsection{Markov State Model} 

First, clustering was performed to discretize the simulated trajectories in terms of $M$ states, i.e. clusters. For this, we used our implementation of the hierarchical vs-CNN algorithm~\ref{appendix:software}. For the BD trajectory, the similarity threshold was fixed to $N=20$ due to the dense sampling of $5\times10^{5}$ frames. Note that the results are rather insensitive to the initially fixed value of $N$ because the hierarchy iterates over decreasing $R_i$, and thus over increasing density thresholds. An initial scan for a cutoff, which clusters 99\% of the data, yielded $R_0=3.0\times10^{-2}~\lambda_B$. Subsequently, the hierarchical approach was initiated with $\Delta F = 1~k_\mathrm{B}T$. Applying Eq.~\eqref{eq:hierarchy} gave the hierarchical levels at $R_1=1.8\times10^{-2}~\lambda_B$, $R_2=1.1\times10^{-2}~\lambda_B$, and $R_3=0.68\times10^{-2}~\lambda_B$. The minimum cluster size to keep and to split was set to $N_\mathrm{keep}=N_\mathrm{split}=1000$. For comparison, the non-hierarchical vs-CNN clustering was also performed at the above mentioned but fixed cutoff values $R_1$, $R_2$, and $R_3$ with $N=20$. In the plain vs-CNN approach, $N_\mathrm{keep}=1000$ was fixed.

The six-dimensional PC data of the villin headpiece was hierarchically clustered with fixed $N=5$. First, the initial $R_0 = 1.0$ was found to cluster 99\% of the data. Then, the hierarchical vs-CNN was performed on all pairs of $\Delta F = \{0.2, 0.4, 0.8\}$ and $N_\mathrm{keep}=\{10, 100, 1000\}$. Additionally, the minimal size of a cluster to split was picked from $N_\mathrm{split}=\{100, 1000, 10000\}$, while $N_\mathrm{keep} < N_\mathrm{split}$ was ensured. To reduce memory usage during hierarchical vs-CNN clustering of the data using 15, 20, and 30 PCs every second frame was used such that the time resolution of the trajectory was reduced to 400~ps. The value $N=3$ was set such that $R_0 = \{3.2, 4.2, 5.5\}$ were found to cluster 99\% of the data with 15, 20, and 30 PCs, respectively. Also, the parameters $\Delta F = 0.4~k_\mathrm{B}T$, $N_\mathrm{split}=5000$, and $N_\mathrm{split}=500$ were used in these cases of higher dimensional PC spaces.

Next, the discretized trajectories were used for the milestoning approach in the maximum likelihood estimator for MSM construction of the PyEMMA software package~\cite{PyEMMA2015}
The sampled transitions between the $M$ clustering states determine a $M\times M$ transition probability matrix $P(\tau)$, where the lag time $\tau$ is the time window by which the transitions are sampled from the trajectories. 
The eigenvalues $\lambda_i$ and eigenvectors $\vec{v}_i$ of $P$ provide the mapping into a Markov jump process. The first eigenvector $\vec{v}_1$ represents the stationary probability distribution on the $M$-dimensional state space and corresponds to the trivial eigenvalue $\lambda_1=1$. Thus, the components of the first eigenvector sum to one and it remains constant upon propagation by $P(\tau)$. The remaining eigenvectors $\vec{v}_2, \dots, \vec{v}_M$ quantify the dynamic modes or the probability flux on the $M$-dimensional state space. Their components are positive and negative such that they sum to zero. The corresponding eigenvalues $\lambda_2, \dots, \lambda_M$ are smaller than one and determine the relaxation timescales,
\begin{equation} \label{eq:MSM_kin}
t_i = -\frac{\tau}{\ln(\lambda_i)},
\end{equation}
with $i>1$. Further details on theory and application of MSMs can be found in Refs~\cite{Chodera2007, Pande2010, Prinz2011, Bowman2013, Chodera2014, Husic2018} and references therein.

\section{Results \& Discussion}

\subsection{Five-well Potential}
\label{sec:results:bd}

\begin{figure}[t]
\centering\includegraphics{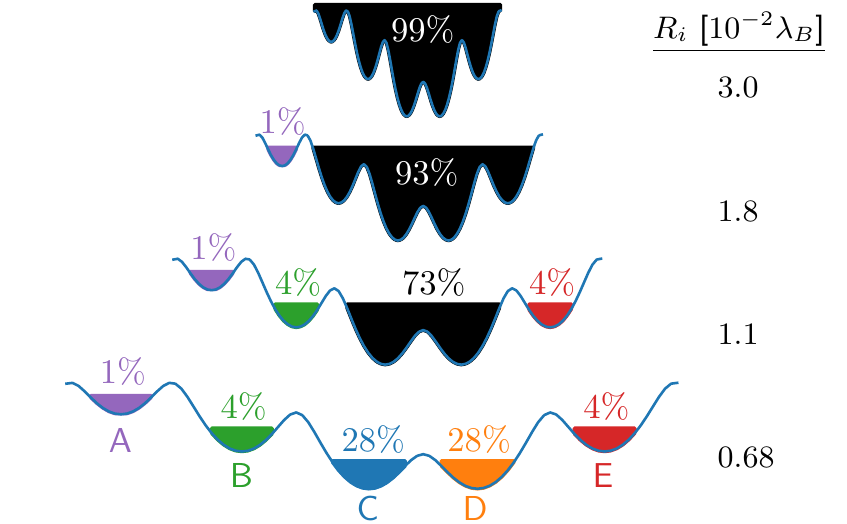}
\caption{Schematic illustration of the hierarchical tree generated by the vs-CNN approach for the circular five-well potential sampled with a Brownian dynamics trajectory. The potential is redrawn at every hierarchical level and the potential is unfolded with broken periodicity between minima A and E. Clusters are indicated with colors. The relative cluster size is denoted in percentage. The corresponding values for $R_i$ are given on the right.}
\label{fig3}
\end{figure}

The vs-CNN clustering algorithm with and without hierarchical scheme was compared on the trajectory of a Brownian particle in a circular five-well potential (Figure \ref{fig2}). 
The hierarchical tree in the five-well potential is shown in Figure~\ref{fig3}, where the circular periodicity of the potential is broken for a one-dimensional representation. The initial tree level at $R_0=3.0\times 10^{-2}\lambda_B$ clusters $99\%$ of the data in a single cluster. Decreasing the cutoff to $R_1$ separates a cluster in the highest energy basin A (purple), which comprises around $1\%$ of the data points. At $R_2$, two clusters of a relative size of $4\%$ in the two energy basins B (green) and E (red) are generated. At the lowest cutoff $R_3$~=~$0.68\times10^{-2}~\lambda_B$, two clusters comprising $28\%$ each are identified in the lowest energy basins C (blue) and D (orange). 

\begin{figure}[t]
\centering\includegraphics{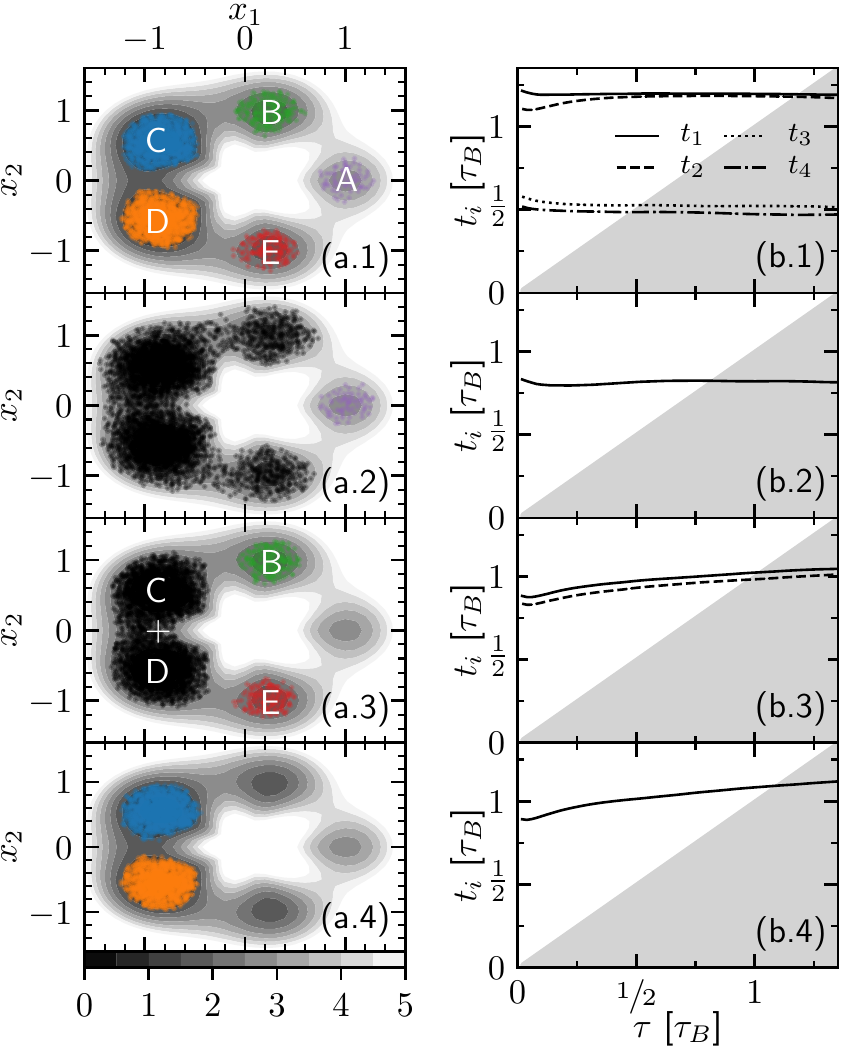}
\caption{Clusters in the two-dimensional potential (a) and implied timescales of the MSMs based on these clusters (b) from a Brownian dynamics trajectory. Results are shown for the hierarchical vs-CNN approach starting at $R_0=3.0\times10^{-2}~\lambda_B$ (1), as well as the non-hierarchical vs-CNN approach with $R$ set to $1.8\times10^{-2}~\lambda_B$ (2), $1.1\times10^{-2}~\lambda_B$ (3), and $0.68\times10^{-2}~\lambda_B$ (4). $N$~=~20 was used. The intersection of the smallest implied timescale with the gray area marks the upper bound for $\tau$ such that the dynamics is properly sampled.}
\label{fig4}
\end{figure}

Figure~\ref{fig4}~a.1 shows the corresponding five clusters from the hierarchical vs-CNN approach in the two-dimensional potential. The same results with fixed cutoff values $R=\{1.8, 1.1, 0.68\}\times10^{-2}~\lambda_B$ are provided in Figure~\ref{fig4}~a.2 -- a.4. It shows clearly that it is not possible to distinguish all five minima for this simple example using a single cutoff value. Only the hierarchical approach is able to identify the clusters corresponding to all five energy basins.

Next, we built the respective MSMs for the described discretizations above. The implied timescales against the MSM lag time $\tau$ are shown in Figure~\ref{fig4} b. The four timescales of the system with five energy basins can only be resolved with the five clusters from the hierarchical vs-CNN approach (Figure~\ref{fig4} b.1). Two timescales are clearly above $1~\tau_B$, while the other two timescales are around 0.5~$\tau_B$. As the non-hierarchical discretizations consist of only two or three clusters, the respective MSMs contain likewise only one or two implied timescales (Figure~\ref{fig4} b.2 -- b.4). The model (2) with one cluster in the highest energy basin A and one cluster spanning across the energy basins B--E predicts a timescale around 0.75~$\tau_B$. Although the implied timescale is converged for all MSM lag times $\tau$, it is clearly off from the true timescales. In contrast, the implied timescales of the two models (3) and (4), which neglect the highest energy basin A, do not converge and show a monotonic dependence on $\tau$. 

\begin{figure}[t]
\centering\includegraphics{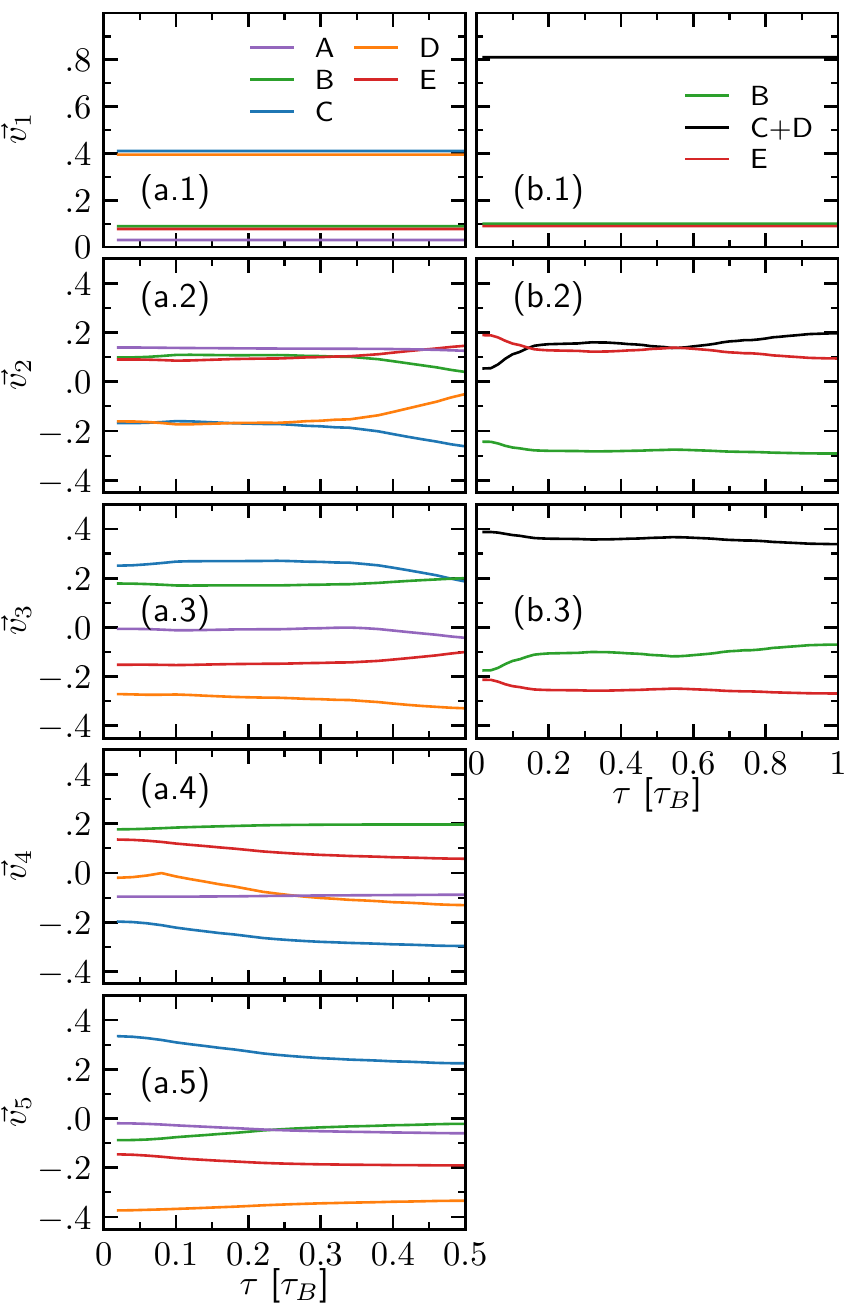}
\caption{Eigenvector components as a function of the MSM lag time $\tau$ for the hierarchical vs-CNN approach (a) and the non-hierarchical vs-CNN approach with $R=1.1\times10^{-2}\tau_B$ (b). $N$~=~20 was used. The components are shown for the stationary distribution $\vec{v}_1$ (1), the first eigenvector $\vec{v}_2$ (i.e. the slowest process) (3), and the second eigenvector $\vec{v}_3$. For the hierarchical vs-CNN approach, also the last two eigenvectors $\vec{v}_4$ (4) and $\vec{v}_5$ (5) are shown.}
\label{fig5}
\end{figure}

Further, we compared the eigenvectors of the MSMs with more than two clusters (i.e. corresponding to the clusters in Figure~\ref{fig4}~a.1 and a.3). The components of the eigenvectors $\vec{v}_1$ are plotted against the MSM lag time in Figure~\ref{fig5}~a.1 and b.1. The prediction of the stationary distribution is constant across all lag times in both models. Moreover, both models determine that a probability of 0.8 lies in the two lowest energy basins C and D. The remaining probability of 0.2 is distributed over the other basins. Only the MSM from the hierarchical vs-CNN approach is able to resolve the small probability of less than 5\% in the highest energy basin A.

In the case of the hierarchical discretization, the eigenvector $\vec{v}_2$ in Figure~\ref{fig5}~a.2 represents the probability flux along the $x_1$ direction from the clusters in the lowest energy basins C and D to the basins A, B, and E. The eigenvector $\vec{v}_3$ plotted in Figure~\ref{fig5}~a.3 resolves the probability flux across the mirror symmetry of the potential, i.e. along the $x_2$ direction. Thus, the clusters in basins D and E have negative eigenvector components while clusters B and C associate to positive components. The component of the highest energy basin A is zero. The last two eigenvectors $\vec{v}_4$ and $\vec{v}_5$ represent the fluxes from diagonal dynamics that are not aligned with either coordinate axis.

If the discretization is performed without the hierarchical scheme (model from Figure~\ref{fig4}~a.3), the eigenvector $\vec{v}_2$ points from the energy basin B to the clusters in basin E and the basins C+D (Figure~\ref{fig5}~b.2). The second eigenvector in Figure~\ref{fig5}~b.3 points from basins B and E into the large cluster merging energy basins C+D. Thus, the separate eigenvectors represent diagonal probability flux. Taken together, the monotonic increase of the implied timescales in Figure~\ref{fig4}~b.3 and the eigenvectors diagonal to the symmetry axis of the potential in Figure~\ref{fig5}~b.2 and b.3 suggest that a mixing of the originally four timescales and processes occurs. 

This means that a non-hierarchical density-based clustering approach can yield an indecisive timescale interpretation and corrupted dynamic processes. A hierarchical approach increases the robustness of the core-set MSM construction when the energy landscape is generally rugged across different (free) energy levels. 

Note, that for this simple two-dimensional five-well potential we found 35\% noise (Figure~\ref{fig3}) after the hierarchical clustering approach, which led to a proper identification of core sets in the potential minima as can be seen in Figure~\ref{fig4}. In the following section, we elaborate on the interpretation of the noise level and how it can be used for tuning the clustering results.

\subsection{Villin Headpiece}
\label{sec:results:hp35}

Generally, the (vs-)CNN clustering approach yields unassigned noise points. This poses an advantage over full-partitioning clustering algorithms to identify density peaks, i.e. energy basins for core-set MSM construction. Hence, fast crossings of cluster boundaries that are in the ballistic regime are reduced to a minimum.

\begin{figure}[t]
\centering\includegraphics{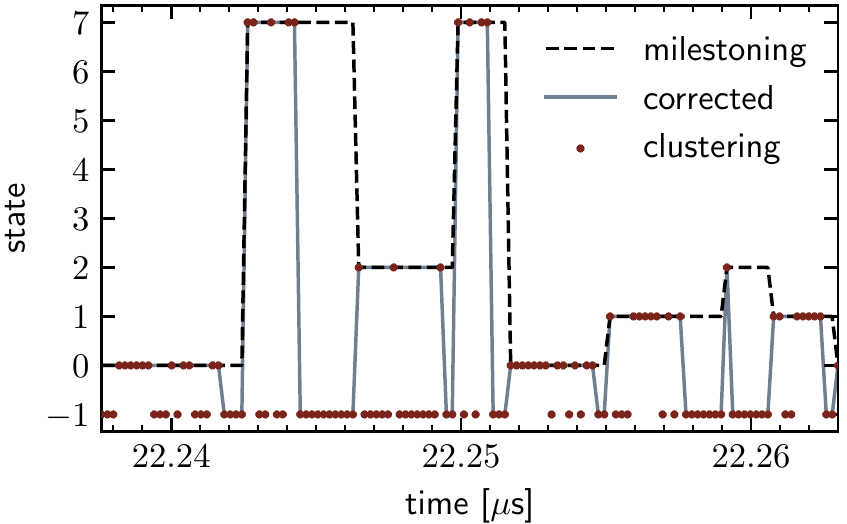}
\caption{Example section of the discretized trajectory of the villin headpiece between 22.24 and 22.26 ns. Noise points are assigned to -1. The trajectory after clustering (red dots) contains many noise points that return into previous state. After correcting for noise points that leave the state unchanged (gray line) only noise points involved in a transition are remain. After milestoning (black dashed line) all noise points have been assigned the previously exited state.}
\label{fig6}
\end{figure}

In the case of a rugged multidimensional space as with the villin headpiece, we find even higher noise levels. However, the noise level can be used as a handle to choose the clustering parameters. One can correct the noise points in a trajectory section that previously visited a given state $X$ and will next return to $X$ again. These noise points, which are not part of a state transition in the time series, can therefore be assigned to state $X$. This correction scheme is illustrated for part of the trajectory of the villin headpiece in Figure~\ref{fig6}. The majority of frames of the discretized trajectory after clustering (red dots) are noise points, i.e. the state is -1. When correcting for such noise points, only state transitioning noise frames remain (gray line). The remaining noise in the corrected trajectory quantifies the actual time spent in transition regions. Furthermore, the milestoning trajectory that is obtained for MSM construction (black dashed line) assigns all noise points to the previously visited cluster.

\begin{figure}[t]
\centering\includegraphics{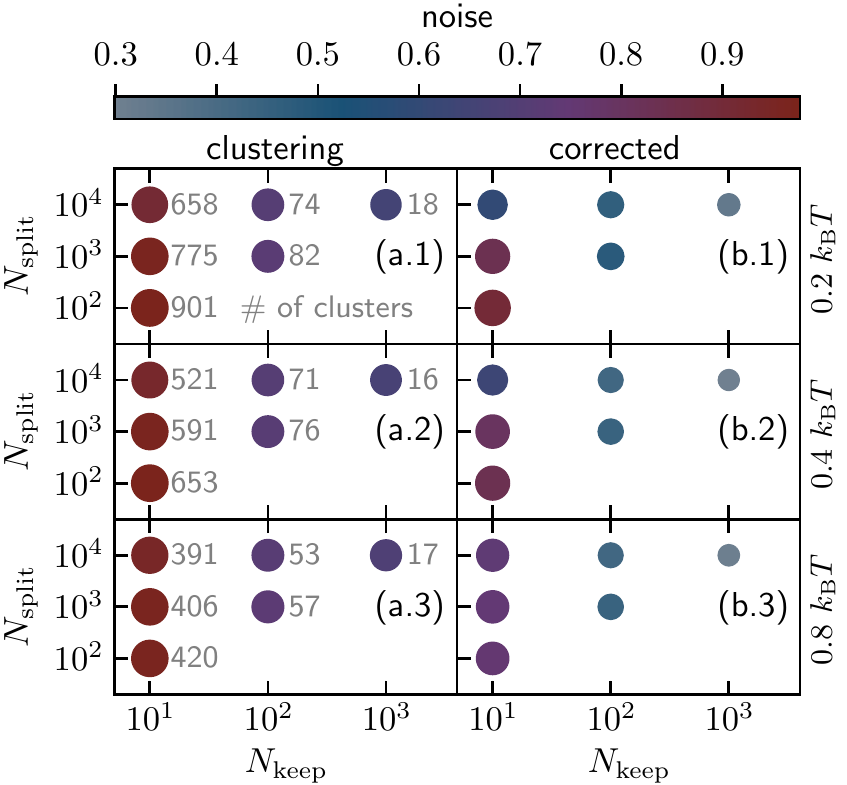}
\caption{Schematic illustration of the noise levels after clustering and after correcting for non-transitioning noise points for varying $N_\mathrm{split}$ and $N_\mathrm{keep}$ values. Panels~a.1-a.3 show for $\Delta F = \{0.2, 0.4, 0.8\}~k_\mathrm{B}T$ the relative amount of noise after clustering and the number of clusters (gray). Panels~b.1-b.3 show the corrected relative amount of noise for different $\Delta F$.}
\label{fig7}
\end{figure}

This means that correcting the relative amount of noise points for frames, which are not involved in a transition, gives a way to decide on a particular parameter choice of $N_\mathrm{keep}$ and $N_\mathrm{split}$. Figure~\ref{fig7} shows the relative amount of noise on the six dimensional PC space for different pairs of $N_\mathrm{split}$ and $N_\mathrm{keep}$ for each value of $\Delta F = \{0.2, 0.4, 0.8\}~k_\mathrm{B}T$ after clustering and after correcting for non-transitioning noise points. Additionally, the number of clusters are tabulated along the respective parameter setting. Figure~\ref{fig7}~a.1-a.3 show that a (too) small value of $N_\mathrm{keep}$ (i.e. 10) leads to very high noise levels of more than 90\% and the number of clusters ranges from 391 to 901 after clustering. After correcting for non-transitioning noise, all noise levels remain above 50\%. If $N_\mathrm{keep}$ is increased to 100, the noise level after clustering is more than 70\% while the corrected noise level is below 50\%. In this case, the amount of clusters ranging from 53 to 82 is comparable to the number of clusters for the same villin headpiece data using the Sittel-Stock clustering~\cite{Nagel2019}. Increasing $N_\mathrm{keep}$ to 1000 reduces the corrected noise levels even further to less than 40\%. Overall, the number of clusters and noise levels depend mainly on the value of $N_\mathrm{keep}$, while $N_\mathrm{split}$ and $\Delta F$ have relatively little influence. Only, for $N_\mathrm{keep} = 10$ and small $\delta F$, a trend toward decreasing corrected noise levels can be observed for increasing $N_\mathrm{split}$. Note that the density-based clustering can still give multiple clusters inside a particular energy basin, i.e. density peak, because fine clustering might resolve density fluctuations from finite sampling. Such fine discretizations are inherited by the original CNN algorithm. Note that the vs-CNN already reduces overdiscretization as exemplified in Appendix~\ref{appendix:CNNvsvs-CNN}. Based on the parameter scan in Figure~\ref{fig7}, one can choose a clustering based on reduced noise levels and a desired coarse or fine resolution of metastable states, i.e. the number of clusters.

\begin{figure}[t]
\centering\includegraphics{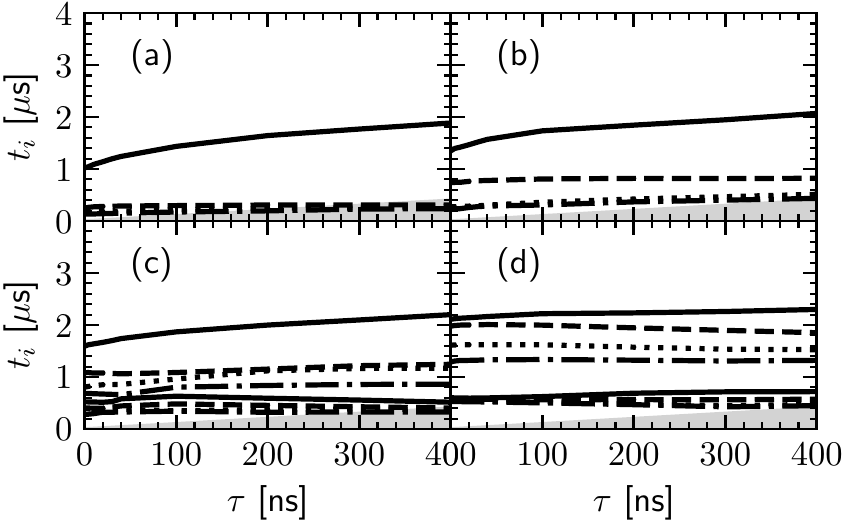}
\caption{Implied time scales against MSM lag time from the discretization on an increasing number of PCs: six PCs as in Ref.~\cite{Nagel2019} (a), the first 15 PCs (b), the first 20 PCs (c), and the first 30 PCs (d).}
\label{fig8}
\end{figure}

We chose the coarser clustering with $\Delta F = 0.4~k_\mathrm{B}T$, $N_\mathrm{keep}=1000$ and $N_\mathrm{split}=10000$ to construct the final MSM. As can be seen in Figure~\ref{fig8}~a, the slowest implied timescale on the discretization on the six PCs is not well converged, i.e. not constant across varying MSM lag times. This is in line with the observations by Nagel et al.~\cite{Nagel2019} using the Sittel-Stock clustering on the same data set. Nagel et al. used dynamical coring~\cite{Jain2014} to partially remove the non-Markovian effects.

Here, we increased the dimensionality of the data (i.e. the number of PCs) to check if an additional correction like dynamical coring is indeed required. The implied timescales of the different discretizations that are larger than 100~ns on the first 15, 20, and 30 PCs are shown in Figure~\ref{fig8}~b-d. Taking into account 15 or 20 dimensions improves the Markovianity of the slowest implied timescale slightly. At the same time more processes are resolved below or roughly around 1~$\mu$s. However, there ist still a lag-time dependence of the slowest implied timescale. Increasing to 30 PC dimensions resolves four processes that are clearly above 1~$\mu$s (Figure~\ref{fig8}d). Most importantly, the implied timescales become mostly independent to the choice of MSM lag time $\tau$, and thus fulfill the Markovianity best even for small $\tau$.

\subsection{General Discussion}

For both the BD system and the villin headpiece, we observed a mixing of slow processes leading to non-constant implied timescales when the multidimensional pathways were not properly resolved in the MSM. In the simple BD example in Figure~\ref{fig4}~a.4 and b.4, we found that the transition along the circular closure and the direct path across the barrier connecting both energy basins got combined in the two-state discretization. In the villin headpiece case, the loss of information due to oversimplified projections of multidimensional data is also obvious. The higher dimensional pathways remain undetected because mutually projected energy basins cannot be distinguish.

This behavior can be understood on a general basis. Markovian and non-Markovian processes can be interconverted by reduction or extension of the dimensionality of the configuration or even phase space, respectively~\cite{Risken, vanKampen}. For instance, a non-Markovian stochastic process with an exponentially decaying memory kernel can be represented by a Markovian process with additional auxiliary variables~\cite{Pollak&Berezhkovskii}. The inverse idea of generating a non-Markovian description when reducing the number of reaction coordinates of a Markovian process is equivalent.

To address the issue in practice, the common procedure has been established to choose the dimensionality based on high levels of cumulative variance after PCA or TICA~\cite{Perez2013, Schwantes2013}.

\section{Conclusion}
\label{sec:conclusion}

Starting from the density-based CNN clustering algorithm, we developed the vs-CNN approach with explicit estimation of the data-point density. The density estimation enables a direct interpretation of a common free energy in terms of the Boltzmann inversion, which in turn led to the intuitive hierarchical vs-CNN algorithm. The hierarchical approach maps the choice of vs-CNN's cutoff and similarity parameters into a single parameter $\Delta F$ for an automatized hierarchical tree search. This algorithm is a specialized scheme for Boltzmann distributed data of dynamic systems such as MD simulation trajectories. The hierarchical vs-CNN clustering represents a robust scheme for MSM construction, which are nowadays an integral tool to understand MD simulations of increasingly complex systems. 

The mechanism of the hierarchical vs-CNN algorithm was illustrated step-by-step using the simple example of a Brownian particle in a circular five-well potential, where the minima are at different energy levels. Such a simple example demonstrated the importance of the hierarchical scheme for density-based clustering even for the qualitative interpretation of the slowest timescales and processes. If energy basins were not resolved by a plain density-based clustering, a mix of timescales seemed to distort the implied timescales and the Markovianity of the models. In contrast, when resolving all basins across the different energy levels with the hierarchical vs-CNN, all pathways and thus timescales have been properly separated in the circular closure of the BD setup.

Further, we tested the hierarchical vs-CNN on a 305~$\mu$s villin headpiece folding simulation from Ref.~\cite{Piana2012}. First, we used the more complex data set to describe a parameter search scheme based on the resulting noise levels and number of clusters. As a result, $N_\mathrm{keep}$ was identified as the main parameter determining the resolution of the discretization. This parameter was inherited from the original CNN algorithm. The sensitivity of the clustering on $\Delta F$ and $N_\mathrm{split}$ was found to be minor. In particular, the choice of $N_\mathrm{split}$ only becomes more relevant for a fine free-energy resolution of the hierarchy, i.e. small $\Delta F$. However, a fine free-energy resolution increases the computational cost of the clustering because the number of hierarchical steps increases with decreasing $\Delta F$. In general, we recommend to set $\Delta F \sim 0.1~k_\mathrm{B}T$. If the computational cost is a limiting factor for a given data set (depending on number of frames, dimensionality, etc.), $\Delta F$ can be increased. As an upper boundary, $\Delta F \lesssim 1~k_\mathrm{B}T$ still detects all states at the boundary of thermal stability but potentially neglects some smaller, fast living states. The parameter $N_\mathrm{split}$ can also be set to a fixed value with $N_\mathrm{split} > 2\cdot N_\mathrm{keep}$. Most importantly, a coarse search for $N_\mathrm{keep}$ across orders of magnitude is recommended, which can especially be used to tune the clustering resolution.

After determining the discretization of the villin headpiece data, we compared MSM results for different data dimensionality. Like Nagel et al.~\cite{Nagel2019}, we observed a dependence of the slowest implied timescale on the MSM lag time when considering six PCs. Nagel et al. used dynamical coring to correct for remaining non-Markovian contributions. We instead found that the behaviour of the implied timescales can be improved by increasing the number of PCs. The best resolution of timescales was observed for clustering on 30 PCs.

We discussed a potential lack of metastable state detection in multidimensional pathways for the villin headpiece example, and compared to the missing path pathway resolution of the BD data. Importantly, reducing the dimensionality too much can result in oversimplified projections of multidimensional data sets, and thus fail to detect energy basins. We recommend therefore to check for proper timescale separation before employing additional algorithmic steps to improve Markovianity, as these may not be necessary.

In summary, insufficient metastable state resolution due to non-hierarchical density-based clustering strategies or strongly reduced dimensionality yields mixing of timescales and their dependence on the MSM lag time. The presented hierarchical vs-CNN algorithm improves timescale separation for core set MSM construction and simplifies the algorithmic workflow.

\begin{acknowledgements}
R.G.W. and S.R. gratefully acknowledge Markus Aebi for his generous support. The authors thank D. E. Shaw Research for providing the villin headpiece trajectory. The authors gratefully acknowledge financial support by the Swiss National Science Foundation (Grant Number 200021-178762) and by ETH Zurich (ETH-34 17-2).
\end{acknowledgements}

\section*{Data Availability}
An open-source implementation of the vs-CNN clustering algorithm is provided, see Appendix \ref{appendix:software}.

\appendix

\section{Clustering Software}
\label{appendix:software}

The open-source code and documentation of our clustering software is distributed and maintained on 
\href{https://github.com/rinikerlab/vsCNN}{https://github.com/rinikerlab/vsCNN}.
The software implements the vs-CNN as well as the CNN algorithms whereas both can be used with fixed density-based thresholds or combined with the free-energy hierarchy. 

As mentioned in the Introduction, (vs-)CNN requires storing all neighborhoods $M_i$, which requires a lot of memory. To reduce the memory usage, our implementation makes use of adjacency lists that are known to improve the storage usage for sparse data sets.

\section{Overdiscretization with CNN}
\label{appendix:CNNvsvs-CNN}

\begin{figure}[t]
\centering\includegraphics{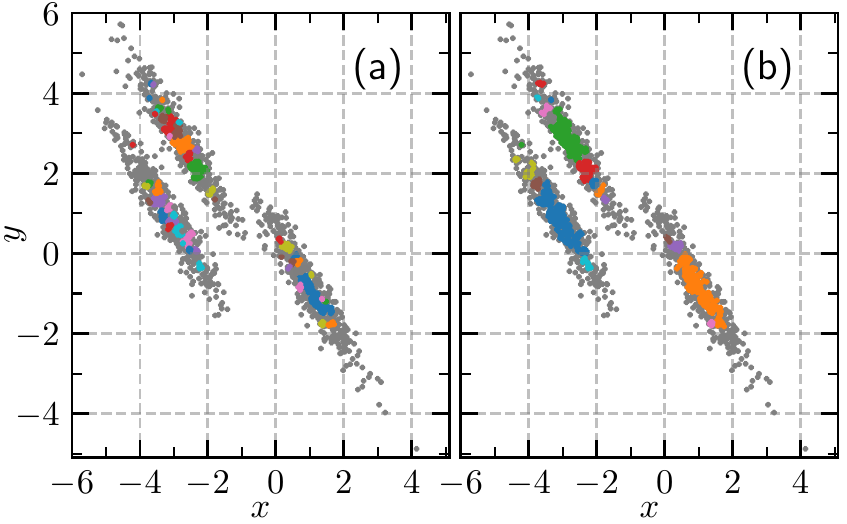}
\caption{Comparison of the CNN (a) and vs-CNN (b) clustering algorithms on a two-dimensional example data set with 1500 points taken from the \textit{scikit-learn} Python package~\cite{scikit-learn}. The thresholds $R=0.1$ and $N=4$ were used. Noise points are shown in gray. Points that belong to the same cluster have the same color, while the number of colors in the palette is repeated.}
\label{fig9}
\end{figure}

We assessed how the density-based clustering changes when the similarity criterion of the CNN algorithm is replaced by Eq.~\eqref{eq:cd} for the vs-CNN algorithm. For this, we used a two dimensional test set with 1500 points from the \textit{scikit-learn} Python package~\cite{scikit-learn}. The data set was clustered using the CNN and vs-CNN algorithms with $R=0.1$ and $N=4$ (while $N_\mathrm{keep}=2$). As can be seen in Figure \ref{fig9}, the CNN algorithm generates an overly fine-grained clustering, which is reduced with the vs-CNN algorithm. vs-CNN creates 19 clusters and removes 48\% of the data as noise. Each of the three largest clusters contains more than 200 points. The fourth largest cluster contains 29 points and the size of the remaining clusters is successively decreasing. Such a gap in cluster size enables removal of the additional clusters and the identification of their possible redundancy through a parameter scan of $N_\mathrm{keep}=30$ to $\approx 200$. In contrast, CNN increases the number of clusters by more than three-fold to 60 and classifies 56\% of the data as noise. Moreover, the clustering does not exhibit a gap in cluster size. The first few cluster sizes are 116, 60, 44, 40, 37, 24 etc., which does not allow for an agnostic identification of the three density peaks. Of course, CNN achieves clear identification of the three dense regions for increased $R$ or decreased $N$. In turn, vs-CNN can similarly generate many more redundant clusters for denser criteria, but the reformulated algorithm clearly reduces the range of overdiscretizing parameter settings. Mechanistically, the vs-CNN algorithm explicitly estimates the point density, which can be larger than the pure cardinality criterion used in the CNN algorithm. Thus, the vs-CNN algorithm still clusters points based on the density, which is larger than $N$, while the cardinality might not be. This leads to fewer `redundant' clusters and less noise.


%

\end{document}